\documentclass[referee]{aa}

\usepackage{graphicx}
\def\ls{{_<\atop^{\sim}}}
\def\gs{{_>\atop^{\sim}}}

\begin{document}
%%%%%%%%%%%%%%
\title{The variable X-ray light curve of GRB 050713A: the case of refreshed
shocks.\footnote{Based on observations made with the Swift NASA satellite and with
XMM-Newton, an ESA science  mission with  instruments and contributions directly funded by
ESA Member States and on observations collected with Telescopio Nazionale Galileo,
 the Liverpool Telescope and the Nordic Optical Telescope}}

\author{D. Guetta\inst{1}, F. Fiore\inst{1}, V. D'Elia\inst{1},
R. Perna\inst{2},
L.A. Antonelli\inst{1}, S. Piranomonte\inst{1}, S. Puccetti\inst{1},
L. Stella\inst{1}, L. Angelini\inst{3}, N. Schartel\inst{4},S.Campana\inst{6},
G. Chincarini\inst{5,6}, S. Covino\inst{6}, G. Tagliaferri\inst{6},
D. Malesani\inst{7},C. Guidorzi\inst{8}, A. Monfardini\inst{8}, C.G. Mundell\inst{8},
de Leon Cruz\inst{9}
A.J. Castro-Tirado\inst{10}, S. Guzly\inst{10}, J. Gorosabel\inst{10},
M. Jelinek\inst{10}, A. Gomboc\inst{11}
}

\institute{
INAF --- Osservatorio Astronomico di Roma, via Frascati 33, I-00040
Monteporzio Catone (Roma), Italy
\email{guetta@mporzio.astro.it}
\and
JILA and Department of Astrophysical and Planetary Sciences, University
of Colorado, 440 UCB, Boulder CO, USA
\and
NASA, Goddard Space Flight Center, Greenbelt, MD, USA
\and
XMM-Newton SOC, ESAC, ESA, Apartado 50727, 28080 Madrid, Spain
\and
Universit\`a degli studi di Milano-Bicocca, Dipartimento di Fisica,
Piazza delle Scienze 3, I-20126 Milano, Italy
\and
INAF --- Osservatorio Astronomico di Brera, via E. Bianchi 46,
I-23807 Merate (Lc), Italy
\and
International School for Advanced Studies (SISSA-ISAS), via Beirut
2-4, I-34014 Trieste, Italy
\and
Astrophysics Research Institute, Liverpool John Moores University,
Twelver Quays, Birkenhead CH41 1LD, UK
\and
IAC: Instituto de Astrof\'{\i}sica de Canarias (IAC), C/. Via L\'actea, 38200 La Laguna,
Tenerife, Spain.
\and
Instituto de Astrofisica de Andalucia (IAA-CSIC), Apartadode Correos,
Granada, Spain.
\and
A. Gomboc
Faculty of Mathematics and Physics, University
of Ljubljana, Jadranska 19, 1000 Ljubljana, Slovenia
}
\date{May, 19 2006}

\abstract{ We present a detailed study of the spectral and temporal
properties of the X-ray and optical emission of GRB050713a up to 0.5
day after the main GRB event. The X-ray light curve exhibits large
amplitude variations with several rebrightenings superposed on the
underlying three-segment broken powerlaw that is often seen in
Swift GRBs. Our time-resolved spectral analysis supports the
interpretation of a long-lived central engine, with rebrightenings
consistent with energy injection in refreshed shocks as slower
shells generated in the central engine prompt phase catch up with
the afterglow shock at later times. Our sparsely-sampled light curve
of the optical afterglow can be fitted with a single power law
without large flares. The optical decay index appears flatter than the
X-ray one, especially at later times.   }

\authorrunning {Guetta et al.}
\titlerunning {GRB 050713A: the case of refreshed shocks}

\maketitle

%%%%%%%

\section{Introduction}

A commonly accepted wisdom in the pre-Swift era was that the optical
and X-ray afterglows of Gamma Ray Bursts (GRBs) showed a smooth power
law decay with time after the burst (e.g. Laursen \& Stanek
2003). This behavior was found to be consistent with the fireball
model, where the afterglow flux is produced when a relativistic blast
wave propagating into an external medium. Under the assumption of a
spherical fireball and a uniform medium, Sari, Piran \& Narayan (1998)
showed that the dependence of the flux on frequency and time can be
represented by several power law segments, $F_{\nu}\propto
\nu^{-\beta}t^{-\alpha}$. Before Swift only a few GRBs showing
deviations from the smooth power law light curve were known, see e.g.
the case of GRB 021004  (Bersier et al. 2002; Matheson et
al. 2002). Its lightcurve was densely sampled in the optical,
allowing detailed modeling. The bumps were interpreted as due to
overdensities in the interstellar medium in which the afterglow is
produced (Lazzati et al. 2002; Nakar et al. 2003; Heyl \& Perna
2003). On the other hand, Bjornsson et al. (2004), Nakar et al. (2003)
and de Ugarte Postigo et al. (2005) modeled these fluctuations as due
to several energy injection episodes. Several re-brightenings were
observed also in the optical light curve of GRB 030329 and were
interpreted as due to refreshed shocks (Granot, Nakar \& Piran 2003,
Huang et al. 2006).

This simple picture is now changing since the advent of Swift.
Bright X-ray flares have been recently observed by Swift in almost
half of its detected GRBs (Gehrels et al.  2005; Burrows et al.
2005b; Falcone et al. 2006; Nousek et al. 2005; O'Brien et al.
2006). While some bursts show one distinct flare, like GRB 050406
(Romano et al. 2006), other events like GRB 050502B and GRB 050713A
show several flares (Burrows et al.  2005; O' Brien et al. 2006;
Falcone et al. 2006). One of the main current goals of the GRB
community is to understand the origin of this newly observed light
curve behaviour. Since flares are likely to trace the activity of
the internal engine, they can help us gain a more comprehensive view
of the physical processes governing the early phases of GRB
activity. Flares have been observed in the light curves of both long
and short GRBs.  For the long bursts, King et al. (2005) proposed a
model in which the flares could be produced from the fragmentation
of the collapsing stellar core in a modified hypernova scenario. For
the case of short GRBs, MacFadyen et al. (2005) suggested that the
flares could be the result of the interaction between the GRB
outflow and a non-stellar companion. More recently, Perna et al.
(2006) have analyzed the observational properties of flares in both
long and short bursts, and suggested a common scenario in which
flares are powered by the late-time accretion of fragments of
material produced in the gravitationally unstable outer parts of the
hyperaccreting accretion disk.  Other mechanisms that could produce
flares are of magnetic origin (Gao et al. 2005; Proga \& Zhang
2006). In this paper we concentrate on one particular burst that
presents flaring activity, GRB 050713A, and perform a detailed
spectral and timing analysis of its main flare with the goal of
constraining the physical mechanisms that can be responsible for its
production.

The {\it Swift} BAT localized this burst on 2005, July 13.1866 UT to a
3\arcmin{} radius error circle (Falcone et al. 2005).  The BAT light
curve is characterized by a main event lasting $\sim 13 sec$ that
drops by a factor of 100, followed by two rebrightenings starting
$\sim 53$~s and $\sim 110$~s after a time $t_0$~s, that corresponds to
the time at which BAT started to detect the burst.  The spectrum of
the main event could be fitted with a power law with energy index $
0.58 \pm 0.07$, yielding a fluence of $(9.1 \pm 0.6) \times 10^{-6}$
erg~cm$^{-2}$ (15--350 keV).  The peak flux in 1~s window time is $6.0
\pm 0.4$ ph~cm$^{-2}$~s$^{-1}$ (Golenetskii et al. 2005). Swift slewed
promptly toward the position of the GRB and XRT started to observe
this event just 70~s after the trigger. The XRT light curve was
dominated by a major rebrigthening event 117~s after $t_0$, nearly
coincident with the second rebrightening detected by BAT.  Both the
15--350 keV and the 0.5--10 keV fluxes varied up and down by a factor
of $\sim10$ in about 40~s.  A second XRT rebrightening event occurred
about 186~s after the $t_0$, and other smaller amplitude flares were
seen at later times (around $\sim 10^4$~s). In this paper we focus on
the first major XRT rebrightening event through both a temporal and a
time-resolved spectral analysis. The statistics of the spectra of the
second flare are not good enough to allow a similar detailed analysis.
Furthermore, we use XMM-Newton data to better constrain the late
time afterlow decay.

We show that the rebrightening in the early-time afterglow light curve
is consistent with energy ``injections'', probably due to later shells
that catch up with the main shock at later times.

The redshift of GRB 050713A is not known. The host
galaxy is not detected, and the limit on its magnitude is quite
shallow ($R \ls 23$), due to the high background caused by a nearby
($\sim1$ arcmin), bright star.

\section{Observations and data reduction}

\subsection{Swift XRT observations}

XRT started  observing GRB050713A at 2005-07-13, 04:30:14.9, just 70~s
after the BAT trigger. The observation lasted until 2005-07-13,
08:50:27 (UT), for a total of $\sim 1700$~s net integration time. 
Within this observation, XRT observed the GRB in three consecutive orbits.

The data were reduced using the Swift Software (v. 2.0) and in
particular the XRT software developed at the ASDC and HEASARC (Capalbi
et al.
2005\footnote{http://heasarc.gsfc.nasa.gov/docs/swift/analysis/xrt\_swguide\_v1\_2.pdf}).
Standard screening criteria were adopted to reject ``bad'' events
following Capalbi et al. (2005). In particular, the constraint on the
angular distance between the source position and the satellite
position (which must be less than $0.08$ deg, Capalbi et al. 2005),
led us to exclude the first part of the first orbit and 80\% of the
last orbit. The GRB was observed by the XRT in two observing modes.
During the first satellite orbit the GRB was mostly observed in
windowed timing mode (WT mode, providing 1D imaging). We report here
the WT data only for this orbit.  During the second and third orbits
the GRB was mostly observed in photon counting mode (PC mode,
providing the usual 2D imaging). We report here the PC data only for
these orbits.  We used a $0-12$ grade selection for the PC mode and
$0-2$ for the WT mode. We excluded all the events with energy below
$0.3$ keV, to minimize the background due to the bright Earth limb and
and because the calibration of the data below 0.3 keV are still
uncertain. The intensity of the source was high enough to cause
significant pileup in the PC mode. In order to avoid this pileup we
extracted counts from an annulus with inner radius of $6$ pixels and
outer radius of $20$ pixels (14 and 47 arcsec, respectively).  We then
corrected the observed count rate for the fraction of the XRT Point
Spread Function (PSF) lying outside the extraction region. The
correction was equal to a factor of 3.33 (i.e. 30\% of the PSF lying
inside the annulus). Data in WT mode were not affected by pileup
and therefore we extracted counts from a circular region of $20$
pixels radius (47 arcsec). Physical ancillary response files were
generated with the task {\sc xrtmkarf} to account for the different
extraction regions. For the spectral fits we used the latest
redistribution matrices (version 7).

\subsection{XMM-Newton observations}

XMM-Newton (Jansen et al. 2001) observed this GRB for about 30~ks
starting from 2005-07-13 10:18 UT, about six hours after the trigger,
observation ID 0164571001 (Loiseau et al. 2005).  The data were
processed using the XMM-Newton Science Analysis Survey (SAS)
v.6.1.0\footnote{http://xmm.vilspa.esa.es/external/xmm$_{-}$
sw$_{-}$cal/sas$_{-}$frame.shtml}. We used the raw event files (i.e.
the observation data files, ODF), which were linearized with the
XMM-SAS pipelines, {\sc epchain} and {\sc emchain} for the PN
(Str\H{u}der et al. 2001) and MOS (Turner et al. 2001) cameras
respectively. Events spread at most in two contiguous pixels for the
PN (i.e. grade=0--4) and in four contiguous pixels for the MOS (i.e,
grade=0--12) have been selected. Event files were cleaned from bad
pixels (hot pixels, events out of the field of view, etc.). 
In order to remove periods of high
background, we analyzed the light curves of the counts from the entire
EPIC PN and MOS CCDs at energies higher than 10 keV, where the X-ray
sources contribution is negligible. We rejected time intervals, in
which this count rate was higher than 10 counts s$^{-1}$ and 1.5
counts s$^{-1}$ for the PN and MOS cameras respectively. This
corresponds to rejecting 27\% (15\%) of the PN (MOS) on source time.
The source counts were extracted from a circular region of 47 arcsec
radius (12 pixels).  The background counts were extracted from the
nearest source free region.  The response and ancillary files were
generated by the XMM-SAS tasks, {\sc rmfgen} and {\sc arfgen}
respectively.

\section{Optical observations}

\begin{figure*}
\centering
\includegraphics[width=13cm,angle=-90]{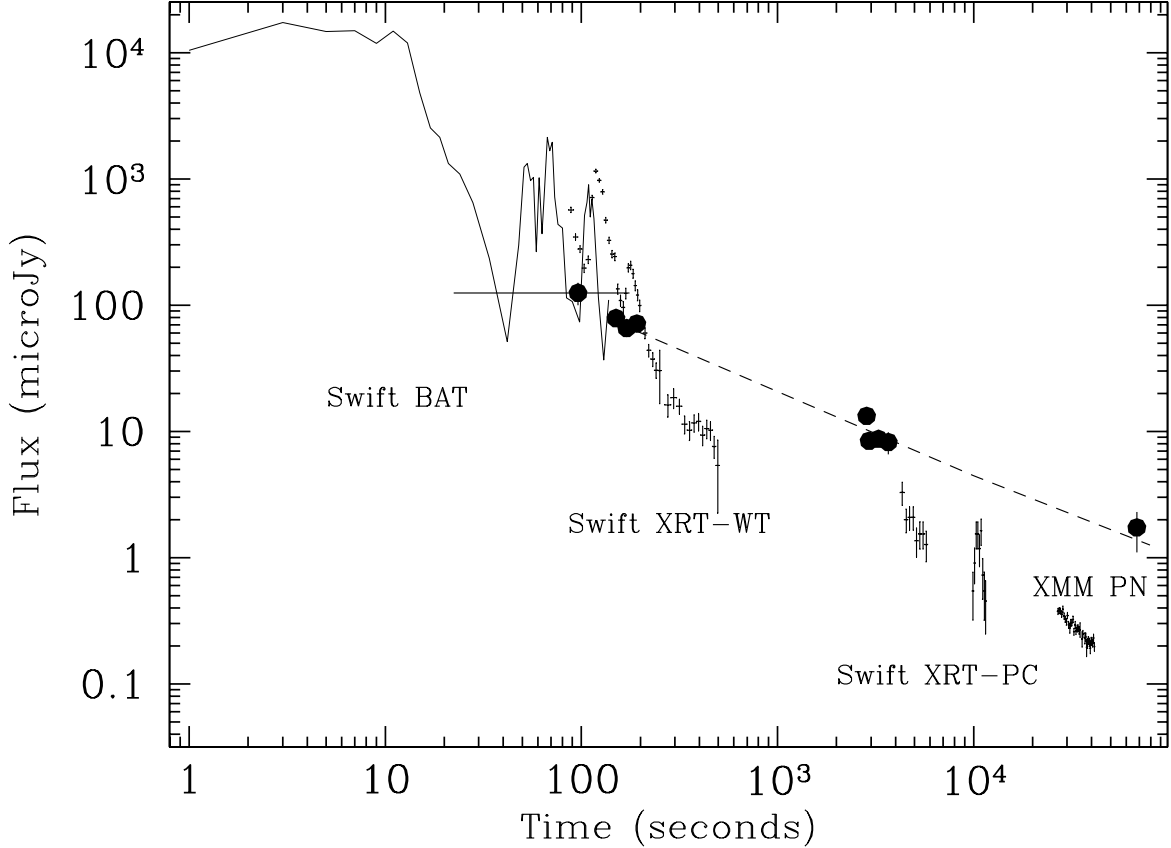}
\caption{\label{lctot} The BAT XRT and XMM-Newton PN  light curve of
GRB 050713A and its afterglow (solid line and points with errors).
The BAT 15--300 keV count rate, the XRT and PN 0.5--10 keV count
rates are converted to a flux at 1 keV assuming a power law spectrum
with average spectral indices $\beta=0.5$ (BAT) and $\beta=1$
(XRT, XMM). Filled circles are the optical observations performed
with the RAPTOR-S, Liverpool 2-m, TNG and NOT telescopes. The dashed
line is the best fit power law decay of the optical data.}
\end{figure*}

UVOT started observing the field at the same time of XRT (75
seconds after the BAT trigger) but did not detect any new source in
the XRT error circle, with a 3$\sigma$ upper limit of V$\sim18$.
Shortly after the receipt of the alert, we started observing the GRB
location in order to look for an optical counterpart.  Observations
were conducted using the Liverpool Telescope (LT), the Telescopio
Nazionale Galileo (TNG), and the Nordic Optical Telescope (NOT), all
located in the Canary Islands. A bright object was discovered inside
the XRT error circle by inspecting the TNG $I$-band images, taken
50~min after the GRB, and was proposed as the GRB afterglow (Malesani
et al. 2005). Its coordinates were $\alpha_{\rm J2000} = 21:22:09.57$,
$\delta_{\rm J2000} = +77:04:29.4$ (0.15\arcsec{} uncertainty). Hearty
et al. (2005) subsequently reported variability for this source,
confirming that it was the GRB counterpart. The robotic 2-m Liverpool
telescope imaged the GRB field starting from 150 seconds after the
burst, detecting the afterglow in the $r\arcmin$ band (Monfardini et
al.  2005). Further imaging was secured at the NOT in the $R$ band
(Jul 13; Guzly et al. 2005) and at the TNG in the $I$ band (Jul
14). Only an upper limit could be obtained at the Observatorio de la
Sierra Nevada (OSN) during the night of Jul 13. Unfortunately,
the GRB position was close ($\approx 1\arcmin$) to a bright foreground
star ($R \sim 6.5$), which caused a bright sky level, so that no
further detection was possible in the following nights.

The TNG and NOT data analysis and reduction were carried out following
the standard procedures. The LT data analysis was performed using the
dedicated LT pipeline (Guidorzi et al. 2006).  Magnitudes were
computed using the SExtractor (Bertin \& Arnouts 1996) and Gaia
software packages. Photometric calibration was obtained by observing
standard fields at the TNG ($I$ band) and OSN ($R$ band).  The
zeropoint, however, was computed at only one epoch, so it may suffer
from systematic uncertainty. LT data were calibrated via Landolt
standards with SDSS calibration (Smith et al.  2002).  We combined all
our data, and included the early RAPTOR $R$-band detection (Wren et
al. 2005). A log of all optical observations is given in Table
1.

Figure \ref{lctot} shows the
light curve of the optical flux at 7000\AA\ obtained by converting
the observed R, r' and I magnitudes into flux at 7000\AA 
assuming a  power law with spectral index of -1.  Considering a
power law flux decay $F(t) \propto t^{-\alpha}$, we found $\alpha
= 0.67 \pm 0.05$. We note that, to within the limitations of the
sampling, the optical light curve appears fairly smooth, with no
signs of the large fluctuations seen in the X-ray region. However,
the coverage is quite scarse, so that no strong conclusion can be
drawn. 

\begin{table}
\caption{Log of Optical observations}
\begin{tabular}{lccccccc}\hline
Mean time &Time since GRB &Exposure time &Instrument  &Filter    &Magnitude      &Flux      \\
(UT)      &(s)            &(s)           &            &          &               &($\mu$Jy) \\
\hline
13.18709  &99.3     &8$\times$10   &RAPTOR      &$R$        &18.4$\pm$0.18  &125$\pm$25   \\
13.22112  &2963     &1$\times$180  &NOT+ALFOSC  &$R$        &21.41$\pm$0.08 &8.4$\pm$0.7  \\
13.22484  &3284     &1$\times$180  &NOT+ALFOSC  &$R$        &21.37$\pm$0.12 &8.7$\pm$1.0  \\
13.22930  &3670     &3$\times$60   &NOT+ALFOSC  &$R$        &21.44$\pm$0.20 &8.2$\pm$1.6  \\
14.08340  &77468    &1$\times$3900 &OSN+CCD     &$R$        &$>$22.50       &$<$3.1       \\
13.18872  &150      &10            &LT+RATCAM     &$r\arcmin$ &19.25$\pm$0.14 &72.4$\pm$9.3 \\
13.18896  &171      &10            &LT+RATCAM      &$r\arcmin$ &19.45$\pm$0.17 &60.3$\pm$9.4 \\
13.18920  &192      &10            &LT+RATCAM      &$r\arcmin$ &19.36$\pm$0.13 &65.5$\pm$7.8 \\
%13.22238  &3059     &2$\times$180  &TNG+DOLORES &$I$        &20.32$\pm$0.07 &19.0$\pm$1.3 \\
13.22238  &3059     &2$\times$180  &TNG+DOLORES &$I$        &20.50$\pm$0.15 &16.1$\pm$1.3 \\
13.98950  &69340    &15$\times$180 &TNG+DOLORES &$I$        &22.70$\pm$0.41 &2.1$\pm$0.8  \\
\hline
\end{tabular}

\end{table}

\section{Results and discussion}

\subsection{Light curves}

Figure \ref{lctot} shows the BAT, XRT and XMM-Newton light curve of
GRB 050713A and its afterglow.  The power law slope of the XRT decay
starting about 200~s after the trigger was $0.67$, very similar to the
optical decay starting 150~s after the trigger (filled circles in
figure \ref{lctot}). The power law decay during the XMM observation is
$1.5$. Thus, the light curve is broadly consistent with that of
many Swift afterglows (Chincarini et al. 2005, Tagliaferri et
al. 2005, Nousek et al. 2005), with a steep decay followed by a
shallower decay and finally a transition to a standard $\alpha=1.5$
decay. The break between the shallow and steep decay should 
occur beween $\sim5000$~s and $\sim20.000$~s otherwise the 
extrapolation of the flux, based on the XMM decay index, would largely 
exceed the flux actually detected by XRT in PC mode.

\begin{figure}
\centering
\includegraphics[width=8cm,height=8cm]{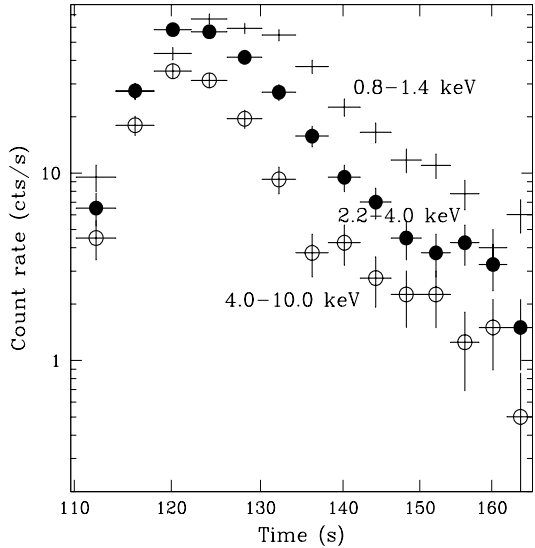}
\caption{\label{picco3}
Light curves of the first flare in three energy bands.}
\end{figure}

\begin{figure}
\centering
\includegraphics{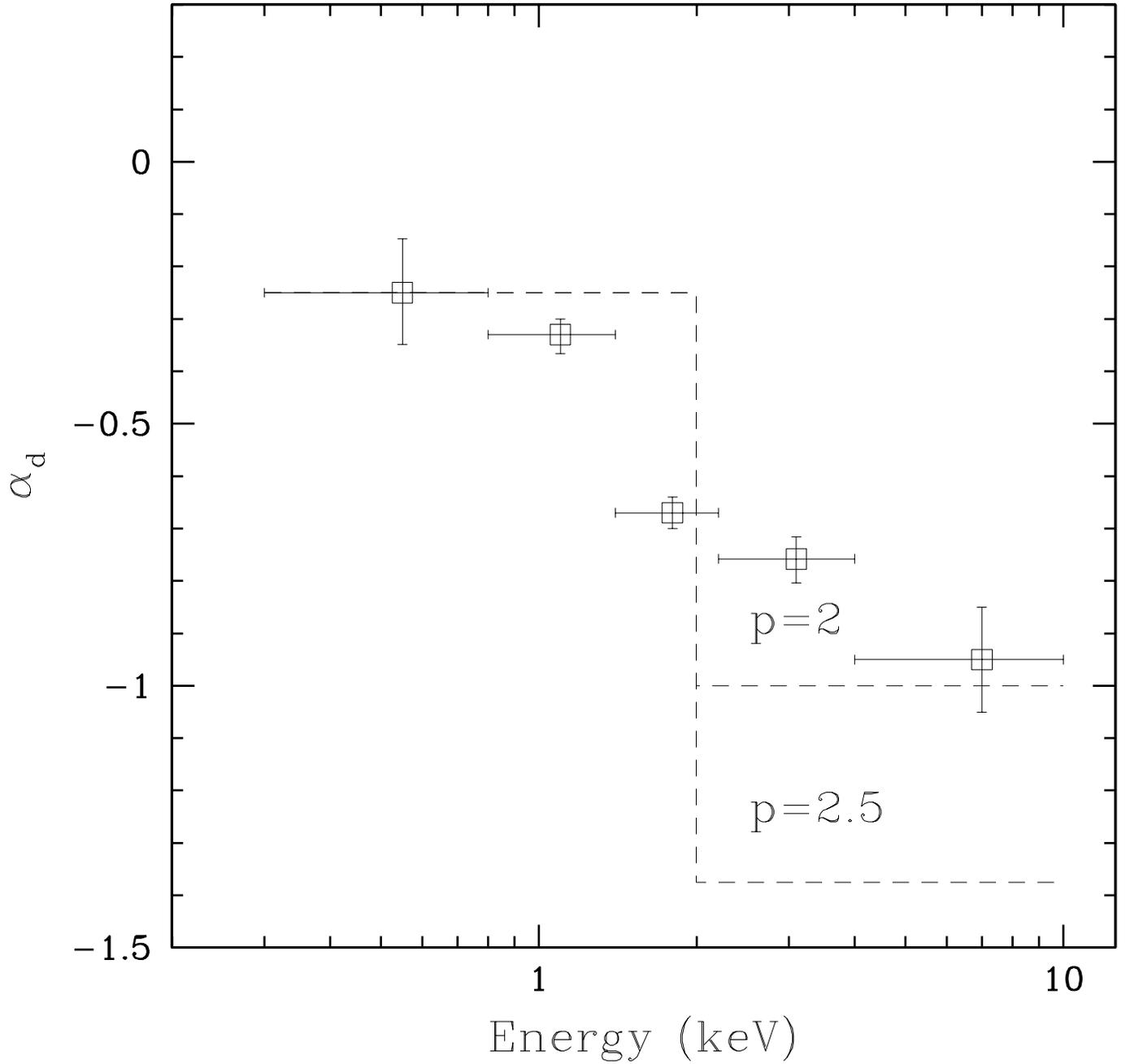}
\caption{\label{discesa} Power law decay indices of the light curves
in the 0.3--0.8, 0.8--1.4, 1.4--2.2, 2.2--4, and 4--10 keV bands,
starting from 108~s after $t_0$, as a function of the energy. The
prediction of the synchrotron model is represented by the dashed lines
considering a broken power law  for two different values of $p$.}
\end{figure}

The BAT data are characterized by two rebrightenings
$\sim 53$~s and $\sim 110$~s after $t_0$.
The early XRT-WT data are characterized by prominent flares
$\sim128$ and 188 seconds after $t_0$.
We note that the first XRT flare overlaps with the
second BAT flare, and that the peak in the hard X-ray band is
shifted of $\sim 10$~s respect to the soft one. An additional flare
 is clearly visible in the XRT PC data.  In the following, we
discuss in more detail these features. Figure \ref{picco3} shows a
zoom of the light curve of the first flare in three energy bands:
0.8--1.4, 2.2--4 and 4--10 keV. Three features are apparent from
this figure: (a) the rise time looks similar in all energy bands;
(b) the decay of the first flare is sharper at higher energies; (c)
the peak of the softer X-ray light curve is shifted by about 10~s
with respect to that of the harder X-ray light curve.  To make these
statements more quantitative, we computed the power law rise indices
$\alpha_r$ from 110 to 118~s after $t_0$ and the power law decay
indices $\alpha_d$ from 124 to 156~s after $t_0$, for the light
curves in the 0.3--0.8, 0.8--1.4, 1.4--2.2, 2.2--4 and 4--10 keV
bands. We used bins of 4~s for the first band and bins of 2~s for
the other 4 bands. The rise time indices were all consistent with
the value 0.6. Conversely, figure \ref{discesa} shows the best fit
decay indices as a function of the energy.  We  note that the power
law rise and decay indices depend strongly on the initial counting
time, $T_0$. Our choice was to set $T_0$ at the beginning of the
flare rise for $\alpha_r$ and at the beginning of the decay for
$\alpha_d$, respectively. Setting $T_0$ at the GRB trigger time
would produce much steeper indices (e.g. $\alpha_d > 3$), possibly
requiring a different physical interpretation of the event,  such as
late internal shocks (Zhang et al. 2005). The underlying decay of
the X-ray light curve between 70 and 200 seconds after the main GRB
events is not well defined in this case, but it appears roughly
consistent with the $\alpha \sim 3$ decay of Zhang et al. (2005).

\begin{figure}
\centering
\includegraphics[width=8cm,height=8cm]{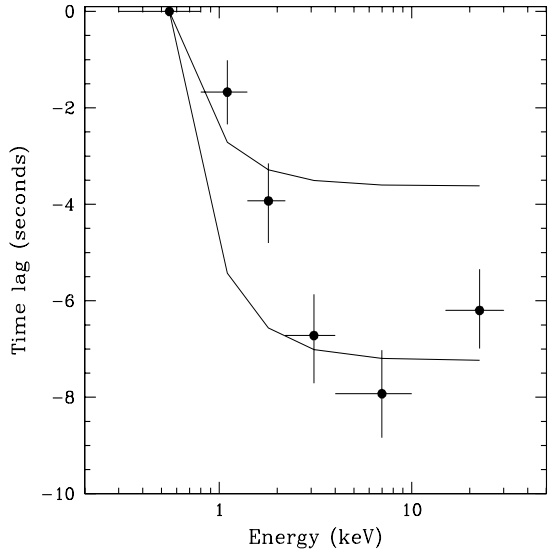}
\caption{\label{ccf} The time-lag computed from the light curves of
the first 128~s of XRT-WT observations, as a function of the
energy. The two solid curves
show the expected time lags for $\epsilon_B^{-3} E_{52}^{-1}
n_{1}^{-2}=10^{-5}$ (upper curve) and $2\times 10^{-5}$ (lower curve) of the 
standard Sari \& Piran (1999) afterglow model.}
\end{figure}

We also computed cross-correlation functions between the light curves
in the four hardest bands and in the 10--30 keV band with respect to
the 0.3--0.8 keV band.  We used the first 128~s of WT data for this
analysis, which includes the two prominent peaks, in 4~s bins.  We
fitted the peak of the resulting cross-correlation functions with a
Gaussian to estimate time lags.  Figure \ref{ccf} shows the time lag
as a function of the energy.   We have also tried to compute the
cross-correlation functions after subtracting a power-law trend,
computed using the first 20~s of WT data and extrapolating the best
fit to the rest of the WT data.  The time lags obtained subtracting or
not subtracting the power law trend are fully consistent one with the
other.

%\begin{figure}
%\centering
%\includegraphics[width=6.5cm,angle=-90]{picco.ps}
%\caption{\label{picco3}
%Light curves of the first flare in three energy bands.}
%\end{figure}

\subsection{Time-resolved spectroscopy}

\begin{figure}
\centering
\includegraphics[width=8cm]{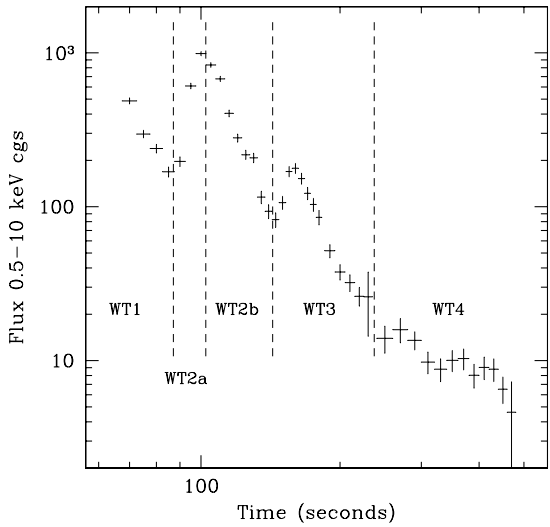}
\caption{\label{lcwt} Light Curve of the first 200 seconds of XRT observations.
Labels mark the time segments in which the five spectra were extracted.
}
\end{figure}

The analysis of the WT light curves of GRB 050713A revealed complex
spectral variations. To investigate  the nature of these variations
we performed a time resolved spectral analysis. We used for this
analysis 5 XRT-WT spectra, selected as illustrated in figure
\ref{lcwt}, 2 XRT PC spectra, corresponding to the two Swift orbits
where XRT operated in PC mode (see Fig. \ref{lctot}), and PN and MOS
spectra covering the full XMM-Newton observations. Results from
XMM-Newton observations were also given by De Luca et al. (2005).

\begin{figure}
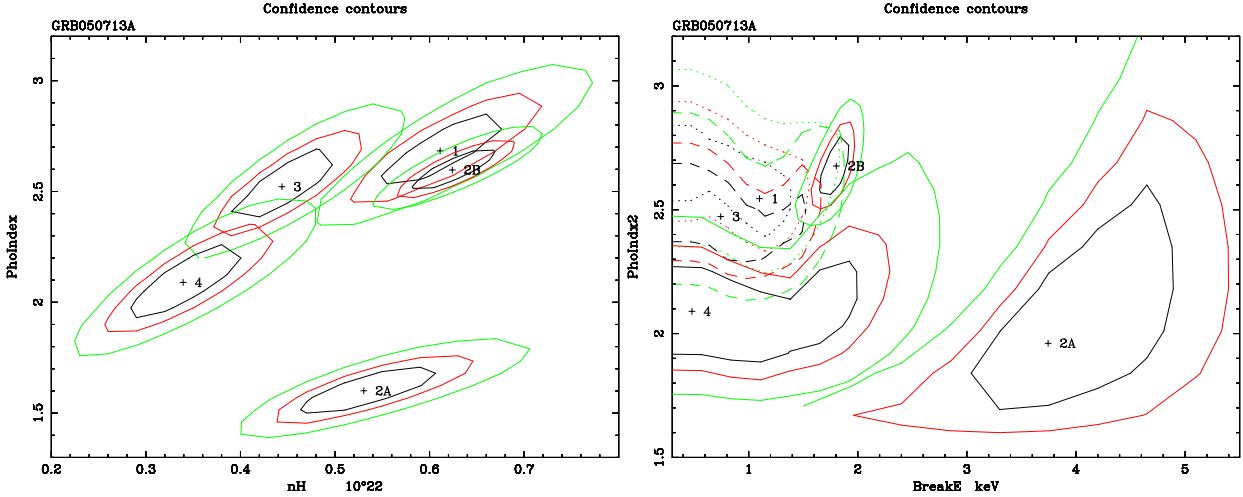

\centering
\begin{tabular}{cc}
\includegraphics[width=6.5cm,angle=-90]{wt_cont_new.ps}
\includegraphics[width=6.5cm,angle=-90]{wt_ge_cont.ps}
\end{tabular}
\caption{\label{cont} a) Left panel: single power law $\Gamma-N_{\rm
H}$ $\chi^2$ confidence contours for the five WT spectra; b) right
panel: broken power law $\Gamma_2-E_{\rm break}$ $\chi^2$ confidence
contours for the five WT spectra.}
\end{figure}

We first fitted the spectra with a simple power law combined with
photoelectric absorption.  We indicate with $\Gamma=\beta+1$ the
photon index. The Galactic column density along the line of sight
to GRB 050713A is $N_H=1.1\times10^{21}$ cm$^{-2}$ (Dickey \&
Lockman 1990). The results of our fits are shown in Table 2.  Large
spectral variations are evident, in particular between the spectra
corresponding to the rise and  fall of the first peak. Figure
\ref{cont} shows the $\chi^2$ contours in the $\Gamma-N_{\rm H}$
plane for the five spectra in WT mode. The power law slope of the
spectra WT2a and WT2b differ by $\Delta\Gamma\approx1$. It is also
interesting to note that, during the decay WT2b, the spectrum
recovers the same spectral slope as during the first decay
WT1.
 This seems to indicate that the first decay WT1 is not the soft
X-ray counterpart of the prompt GRB emission but it is actually the
tail of a flare coincident with the flare detected by BAT in the
15--100 keV band $\sim 53$~s after the $t_0$. In fact, as we see
from Table 2, the spectrum of WT1 is softer than the typical GRB
spectrum. A gradual variation of the spectral power law index
is also evident from WT1 to WT4, where the spectral index is
similar to that observed at later times by XRT in PC mode and by
XMM-Newton.

Motivated by the synchrotron emission model (Sari, Narayan \&
Piran 1998), we then fitted the spectra using a broken power law
model.  We kept the low energy spectral power law index fixed at
$\beta_1=\Gamma_1 -1=0.5$, as expected from standard synchrotron
models and left the high energy power law index
$\beta_2=\Gamma_2-1$ as a free parameter. The results are shown in
Table 3. The $\chi^2$ of spectra WT1, WT3, and WT4 are
indistinguishable from those of Table 2. For spectra WT2a and WT2b the
improvement in $\chi^2$ is significant at the 2\% and 0.15\%
confidence level, respectively, using the F test.

In both sets of fits there is a marginal evidence (between 2 and 3 $\sigma$)
that the absorbing column at $10^4 - 4\times10^4$ seconds after the
trigger was smaller than during the first 100-150 seconds.

\begin{table}[ht!]
\caption{\bf Single power law fits}
\begin{tabular}{lcccc}
\hline
\hline
Spectrum & N$_H$              & $\Gamma$      &   $\chi^2$(dof) \\
         & $10^{22}$cm$^{-2}$ &               & \\
\hline
WT1      & 0.61$\pm$0.10      & 2.68$\pm$0.23 & 51.9(45) \\
WT2a     & 0.53$\pm$0.09      & 1.60$\pm$0.15 & 76.4(69) \\
WT2b     & 0.62$\pm$0.05      & 2.60$\pm$0.15 & 156 (123) \\
WT3      & 0.44$\pm$0.08      & 2.52$\pm$0.22 & 53.1(42) \\
WT4      & 0.34$\pm$0.08      & 2.09$\pm$0.24 & 20.1(34) \\
PC1      & 0.30$\pm$0.16      & 1.66$\pm$0.50 & 5.9(4) \\
PC2      & 0.29$\pm$0.18      & 2.09$\pm$0.40 & 7.2(7) \\
PC1+PC2  & 0.28$\pm$0.08      & 1.75$\pm$0.22 & 31.9(14)\\
XMM      & 0.31$\pm$0.02      & 2.06$\pm$0.04 & 749(715) \\

\hline
\end{tabular}

\end{table}

\begin{table}[ht!]
\caption{\bf Broken  power law fits}
\begin{tabular}{lcccc}
\hline
\hline
Spectrum & N$_H$              & $\Gamma_2$  & E${\rm_break}$ &   $\chi^2$(dof) \\
         & $10^{22}$cm$^{-2}$ &               &  keV        & \\
\hline
WT1      & 0.48$\pm$0.10      & 2.54$\pm$0.20 & $<1.6$      & 51.3(44) \\
WT2a     & 0.50$\pm$0.09      & 1.95$^{+0.75}_{-0.35}$ & 3.75$^{+1.5}_{-1.75}$ & 70.7(68) \\
WT2b     & 0.62$\pm$0.05      & 2.67$\pm$0.17 & 1.8$\pm$0.2 & 144(122)  \\
WT3      & 0.44$\pm$0.08      & 2.47$\pm$0.22 & $<1.8$      & 52.9(41) \\
WT4      & 0.34$\pm$0.08      & 2.10$\pm$0.25 & $<2.3$      & 20.1(33) \\
PC1      &   - & - & - & - \\
PC2      &   - & - & - & - \\
PC1+PC2  &   - & - & - & - \\
XMM      & 0.25$\pm$0.03      & 2.00$\pm$0.10 & $<1.2$      & 745(714) \\

\hline
\end{tabular}

The statistics of the PC spectra is not good enough to
provide robust results adopting models more complex
than a single power law.
\end{table}

\section{Late-time ``energy injection''?  }

Rebrightenings and bumps in the afterglow light curve are generally
due to either density inhomogeneities in the interstellar medium or
to ``energy injections'' at later times. The latter could be
produced by slower shells that catch up with the afterglow shock at
later times. We will consider both possibilities and show that the
energy injection model is favored by the data. Another possibility
is that such flares are produced by late internal shocks, i.e.
slower shells colliding with each other before deceleration (e.g.
Fan \& Wei 2005; Zhang et al. 2005).

First, let us consider the single powerlaw fit to the spectral data.
During the time decay phases (time intervals WT1 and WT2b in
Figure \ref{lcwt}), the single power law spectrum (Table 2) is
consistent with $F_\nu\propto\nu^{-1.6}$. On the other hand, during
the rise phase WT2a, $F_\nu\propto\nu^{-0.5}$  as in the standard
synchrotron model (e.g. Sari, Narayan \& Piran 1998). In this model,
the changes of spectral slopes suggest that during the rise time the
synchrotron frequency $\nu_m$ sweeps across the observation band
(0.5--10 keV).  This is impossible to achieve with a density bump,
since $\nu_m$ is independent of the density for a completely adiabatic
evolution, and almost independent of the density $n$ ($\nu_m\propto
n^{-1/14}$) in a fully radiative evolution. Therefore, we consider the
possibility that the rebrightenings are the result of energy
injections.

A simple argument then shows that the synchrotron frequency is
likely to vary within the 0.5--10 keV band covered by the XRT
observation, and not be outside of the band as implied by the single
powerlaw  fit to the data. Indeed, for a completely adiabatic
evolution, one has $\nu_m\propto E^{0.5}$ (where E is the burst
energy), while for a fully radiative evolution $\nu_m \propto
E^{4/7}$, both implying that the flux rebrightening $F_2/F_1$ scales
with  the frequency change factor as
$\left(\nu_{m,2}/\nu_{m,1}\right)^2$. If the synchrotron frequency
were outside the observation band (i.e. at an energy $\ls 0.5$ keV
before the beginning of the rebrightening, and $\gs 10$ keV after
the end of the rebrigthening), then it would need to vary by a
factor $\gs 20$, implying an energy injection
$\left(\nu_{m,2}/\nu_{m,1}\right)^2\ge 400$.  The rebrightening
implied by this model is much larger than that seen in the data,
$F_2/F_1 \sim 10$, implying that the synchrotron frequency lies
within the 0.5--10 keV band.  For this reason we consider more
physical the broken power law fits in Table 3. The most striking
feature from these fits is that the break frequency is smaller than
1.6 keV in the WT1 spectrum and increases up to several keV in the
 WT2a spectrum (flux rebrightening) and then decreases again at
energies $<$ of 1-2 keV at later times.  This is consistent with the
flat optical to X-ray spectral index from Figure \ref{lctot} at
150-190 seconds after the trigger, indicating that, if  the fast
cooling regime applies, at this time the synchrotron frequency is
between the X-ray and the optical band.

For the same reasons discussed above, the shift of the synchrotron
frequency could not have been caused by a density bump. On the other
hand, in the energy injection scenario, a flux rebrightening by a
factor of 10 would have been produced by an increase in break
frequency by a factor of about 3, fully consistent with the data.

Let us consider now the temporal behavior of the rebrightening as a
function of the energy (sections WT2a and WT2b of the light
curve in figure \ref{lcwt}).  Our analysis shows that the temporal
power law index $\alpha_r$ during the rise time (section WT2a)
is consistent with the constant value $\alpha_r=0.6$ in all the spectral
bands.  During this phase, the temporal evolution is indeed
dominated by the time-dependent energy supply, which is independent
of the frequency band.  The temporal flux decay index  $\alpha_d$
during the decay phase WT2b is shown in Figure \ref{discesa}.  The
lowest and highest energy bands are consistent with straddling the
synchrotron frequency $\nu_m$ for an index $p=2$ of the power-law
distribution of electrons, further confirming the broken power-law
interpretation. Let us clarify this point.  In the low energy band
($<2$ kev), the flux can be written as $F_\nu \propto
(\nu/\nu_c)^{-1/2} F_{\nu,\rm max}\propto t^{\alpha_1}$, while in
the high energy band $F_\nu \propto
(\nu_m/\nu_c)^{-1/2}(\nu/\nu_m)^{-p/2} F_{\nu,max} \propto
t^{\alpha_2}$ where $\alpha_1$ and $\alpha_2$ are the power law
indices of the decay in the low and high energy band respectively
(Sari, Piran \& Narayan 1998). We get a good fit to the data  with
$\alpha_1 = -0.25$ and $\alpha_2 = -3p/4+1/2$. The dashed line in
figure \ref{discesa} shows the results of this model for two values
of $p$, $p=2$ and $p=2.5$.

Taken at face value, the value of p which fits best the decay
indices in figure \ref{discesa} is somewhat different from the value
implied by the $\Gamma$ which fits best the spectrum during the
decay phase WT2b, $p\sim 3$. However it should be considered that
our description of the data is clearly over-simplified, and that
statistical and systematic uncertainties in our determination of
$\alpha_d$ and $\Gamma$ can certainly alleviate this discrepancy.
In conclusion, we find remarkable that, at least qualitatively, the
energy injection model in the framework of the afterglow theory is
able to reproduce both the spectral and the temporal behavior
observed in GRB 050713A.

The time lags between the light curves in the six energy bands
considered here can be further used to constrain the product of the
magnetic field energy fraction $\epsilon_B$, the burst energy, $E$,
and the density of the external medium, $n$, $\epsilon_B^{-3}
E_{52}^{-1} n_{1}^{-2}$, using the expression of the synchrotron
cooling time given by Sari, Piran \& Narayan (1998).  The two curves
in Figure \ref{ccf} show the expected time lags for $\epsilon_B^{-3}
E_{52}^{-1} n_{1}^{-2}=10^{-5}$ and $2\times 10^{-5}$. It is
interesting to note that the synchrotron model does predict a time
lag of the order of 6--9~s between the low energy photons ($<1$ keV)
and higher energy photons (2--10 keV),  and a flattening
of the time lag for energies $>2$ keV as observed.

\section{Summary}

We have presented time and spectral analysis of GRB 050713A, whose
light curve displays several rebrigthenings both in the $\gamma$-ray
and in the X-ray bands. Our time-resolved spectral analysis has
allowed us to conclude that the rebrigthenings are most likely due to
late energy ``injections''. Since the main energy output of the GRB
engine rapidly decays with time (Janiuk et al. 2004), the energy
injections are likely due to either slower shells generated during the
prompt phase of the GRB engine and catching up at later times, or to
later energy production by, for example, disk fragments that accrete
at later times (Perna et al. 2006).

We would like to emphasize two conclusions that came out from the
analysis of the light curve of GRB050713A.  First the behavior of the
time lags between the light curves in the six energy bands is
consistent with what expected from synchrotron cooling.  Second, the
spectral index of the first decay spectrum WT1 is consistent with
that of the decay from the main flare WT2b, thus suggesting that the
WT1 is actually the tail of a flare coincident with the flare detected
by BAT.

The X-ray light curve may present features detected in many Swift
bursts (Chincarini et al. 2005, Nousek et al. 2005, Burrows et al.
2005b). In particular we notice that the XRT decay index starting
200~s after the trigger is flatter than the decay index during the XMM
observation at about 8 hours from the trigger.  This implies that a
break should occur between 5000~s and 20.000~s after the trigger.  The
decay index after the break appears consistent with the predictions of
the uniform ISM afterglow model, while the decay index before the
break is much shallower. This shallow-to-normal decay may be due to
the cessation of the refreshed shock phase. However we cannot exclude
that this is a jet break due to the deceleration of the outflow. The
break occurs at a time when the bulk Lorentz factor of the shock has
slowed to $\Gamma\sim 1/\theta_{jet}$ (Rhoads 1997).

All optical points seem to be typical of the afterglow emission
produced by an external forward shock.  The optical flux varies with
time as $F_{\nu}\sim t^{-0.67}$, which is consistent with the
afterglow decay during the radiative evolution, $F_{\nu}\sim
t^{-4/7}$.  We should notice that this temporal behavior is similar to
the X-ray one between WT4 and XRT-PC.

\bigskip

This research has been partially supported by ASI grant I/R/039/04
and MIUR.
RP acknowledges support from NASA under grant NNG05GH55G, and
from the NSF under grant AST~0507571. We warmly thank the
Telescopio Nazionale Galileo and Liverpool Telescope
teams for precious help in data acquisition.
CG and AG acknowledge their Marie Curie Fellowships from the European
Commission. CGM acknowledges financial support from the Royal Society. AM
acknowledges financial support from PPARC. The Liverpool Telescope is
 operated on the island of La Palma by Liverpool John Moores University at
the Observatorio del Roque de los Muchachos of the Instituto de
Astrofisica de Canarias

\end{document}